\documentclass[twocolumn, prl]{revtex4-1}
\usepackage{graphicx}
\usepackage{epsfig}
\usepackage{color}

\begin{document}
\title{Spin-Orbit Coupled One-Dimensional Fermi Gases with Infinite Repulsion}
\author{Xiaoling Cui$^{\dag \ast}$ and Tin-Lun Ho$^{\ddag\ast}$}
\affiliation{$^{\dag}$Beijing National Laboratory for Condensed Matter Physics,
Institute of Physics, Chinese Academy of Sciences, Beijing 100190, China\\
$^{\ast}$ Institute for Advanced Study, Tsinghua University, Beijing 100084, China\\
$^{\ddag}$ Department of Physics, The Ohio State University, Columbus, OH 43210, USA}
\date{\today}

\begin{abstract}
The current efforts of studying many-body effects with spin-orbit coupling (SOC) using alkali-metal atoms are impeded by the heating effects due to spontaneous emission. Here, we show that even for SOCs too weak to cause any heating, dramatic many-body effects can emerge in a one-dimensional(1D) spin 1/2 Fermi gas provided the interaction is sufficiently repulsive.  For weak repulsion, the effect of a weak SOC (with strength $\Omega$) is perturbative. inducing a weak  spin spiral (with magnitude proportional to $\Omega$). However, as the repulsion $g$ increases beyond a critical value ($g_c\sim 1/\Omega$), the magnitude of the spin spiral rises rapidly to a value of order 1 (independent of $\Omega$). Moreover, near $g=+\infty$,  the spins of neighboring fermions can interfere destructively due to quantum fluctuations of particle motion,  strongly distorting the spin spiral and pulling the spins substantially away from the direction of the local field at various locations.  These effects are consequences of the spin-charge separation in the strongly repulsive limit. They will also occur in other 1D quantum gases with higher spins. 
 \end{abstract}

\maketitle

The recent success of creating spin-orbit coupling (SOC)\cite{Ian_1, Ian_2, Shuai, Jing, Martin} in neutral atoms through Raman processes has stimulated considerable theoretical and experimental activities.  Not only does it lead to new types of Bose condensates\cite{soc-boson}, but also provide opportunities for realizing robust  fermionic pairing\cite{soc-fermion} and  topological matters\cite{topo}.  On the theory side, most studies have  focused on condensate states where mean field descriptions are good approximations. In contrast, there are few studies of spin-orbit effects in strongly correlated states. On the experimental side, there has been significant progress in understanding the properties of spin-orbit coupled Bose condensates\cite{Ian_1, Shuai}, as well as degenerate spin-orbit coupled  Fermi gases\cite{Jing, Martin, Ian_2}. However, it is also found that for alkali metals, the spontaneous emission associated with the Raman process can lead to considerable heating\cite{Ian-heating}, which  prevents one from reaching many novel strongly correlated states that  emerge at very low temperatures. While lowering the  power of the Raman beams  can reduce heating, it will also decrease the strength of spin--orbit coupling, hence losing the effects one sets out to explore. This leads to the crucial question of whether there are pronounced  many-body effects that can be induced by weak spin-orbit couplings.  

In order to produce a large response in the ground state, a large number of excitations must be involved, and the energies of these excitations must be dominated by the perturbation.  Hence, for a SOC with decreasing strength to cause large responses, the ground state of the system must have huge degeneracy. 
Such  states are very rare in many-body systems.  One exception is  the one-dimensional (1D) spin-1/2 Fermi gas with infinite repulsion\cite{Chen, CH, few-body}.   In this case,  fermions irrespective of their spins can not pass through each other. 
The wave function can be  separated into a charge part and a spin part. The former  is given by the wavefunction of spinless fermions,  whereas for the latter,  all spin configurations are degenerate\cite{Chen, CH}. Since the number of spin configurations in a Fermi gas grows exponentially with particle number, the spin degeneracy is huge. 
Thus, no matter how weak the spin-orbit coupling is, there are a large number of excitations lying below the spin-orbit scale that will be strongly affected.

In this paper, we demonstrate the dramatic effect of  the SOC on a strongly repulsive 1D spin-1/2 Fermi gas in a harmonic trap. 
We shall consider the type of SOCs generated in the current Raman scheme\cite{Ian_1, Ian_2, Shuai, Jing, Martin}, which is equivalent to a spatially rotating magnetic field (with wavevector $q$). In particular, we shall consider SOC with energy scale $\Omega$ less than the trap frequency $\omega_{ho}$, 
($\Omega< \omega_{ho}$),  which is weak enough to cause any significant heating. While the SOC will cause the spins to form a spiral, the magnitude and the structure of the spiral changes dramatically as repulsion increases. For example, in the weakly interacting limit, the effect of the SOC is perturbative and the magnitude of the spin spiral is proportional to $\Omega/\omega_{ho}$. However, as the repulsion increases beyond a critical value  $g_{c}\sim 1/\Omega$, the magnitude of the spin spiral (in unit of number density) rises quickly to a value of order of unity (independent of $\Omega$).
The appearance of large spin spiral also greatly reduces the total spin(${\bf S^2}$) of the system. 
 In addition, 
the period of the rotating magnetic field affects  strongly  the interference of neighboring spins. The interference can be so severe that the spins are pulled significantly away from the direction from the local field. 

All these phenomena can be observed in both large samples and small clusters.  They are particularly prominent in small clusters as the charge gap there can be made very large\cite{Jochim}. 
The detection of the sudden increase of the magnitude of the spin spiral, and the quantum fluctuation effects with increasing $q$ will be a demonstration of the strong correlation effects that emerge only  near infinite repulsion.  

Let us consider a 1D  spin-1/2 Fermi gas with repulsive interaction $g>0$ in a harmonic trap in the presence of SOC. The  hamiltonian is 
\begin{equation}
H=\sum_i \left(\frac{p_i^2}{2M}+V(x_i) + V_{i}^{(SO)}\right)+g\sum_{i>j}\delta(x_i-x_j) 
\label{hamil}
\end{equation}
where $V(x)= \frac{1}{2}M\omega_{ho}^2 x^2$ is a harmonic trap with frequency $\omega_{ho}$, and  $ V_{i}^{(SO)}$ is the spin-orbit coupling acting on the $i$-th fermion, 
\begin{eqnarray}
V_{i}^{(SO)} &= - \Omega (e^{iqx_i}S^{-}_{i} + h.c.) = - {\bf B}(x_i)\cdot {\bf S}_{i},  \hspace{0.3in}
 \label{VSO}\\
{\bf B}(x) & =  \Omega \left( \hat{\bf x} {\rm cos}qx  +  \hat{\bf y}  {\rm sin}qx \right) \label{VSO}.   \label{B}  \hspace{1.0in}
\end{eqnarray}
where $\Omega$ is the Raman frequency, and $S^{-}_{i}=S^{x}_{i}-iS^{y}_{i}$. 
What $ V_{i}^{(SO)}$ does is to impart a momentum $q$ to an atom and to  flip its spin. It  is equivalent to the Zeeman energy of a rotating magnetic field ${\bf B}(x)$. 

{\em (A) 1D Fermi Gas at Infinite Repulsion:} For a very weakly interacting Fermi gas in a harmonic trap, (without SOC),  its charge and spin excitations are gapped with energy $~\omega_{ho}$.  
In contrast, a Fermi gas at infinite repulsion has only charge gap ($\omega_{ho}$) and no spin gap. 
For a Fermi gas with $N_{\uparrow}$ spin up and $N_{\downarrow}$ spin down particles, its eigenstates at 
$g= +\infty$  are\cite{Chen,CH}
\begin{equation}
\Psi_{G}(1, 2, ..N)= D(x_1, x_2, ..., x_{N}) \chi^{(S, M)}_{\mu_1, \mu_2, .. \mu_{N}}
\label{GS} \end{equation}
where $x_i$ and $\mu_i$ are position and spin of the $i$-th fermion, 
$D$ is a Slater determinant made up of the eigenstates $\{ u_{n}(x), n=0,1, 2, .. \}$ 
 of the harmonic trap, 
 $\chi^{(S, M)}$ is a spin eigenstate of the total spin $S$ and $S_{z}=M=(N_{\uparrow}-N_{\downarrow})/2$, and 
  $\chi^{(S, M)}$ is a constant within each of the regions $(x_{P1}> x_{P2}>..>x_{PN})$ where $P$ is a permutation of the integers  $ (1,2, ..., N)$.   As a result, the state Eq.(\ref{GS}) satisfies the Schrodinger equation at infinite repulsion. The energy of Eq.(\ref{GS}) is the sum of the energies of the occupied harmonic levels, independent of the spin state. 
{\em In other words, all  energy eigenstates have complete spin degeneracy. All spin excitations are gapless}.  On the other hand, the discreteness of the harmonic oscillator energy levels means the ground state is separated from  the excited state by a charge (or particle) gap $\omega_{ho}$. 
For the SOCs we consider, $\Omega\ll \omega_{ho}$, they will not affect the charge and spin distributions of a  weakly interacting Fermi gas as their excitations are gapped at $\omega_{ho}$. 
However, as $g\rightarrow +\infty$, the spin configuration is strongly affected because of the disappearance of the spin gap. 

{\em  (B) Effective hamiltonian of SOC at $g=+\infty$ and the spin ordered basis:} Let us first focus on the spin state at infinite repulsion $g=+\infty$ and return to the question of the critical interaction $g_c$  for a large spin spiral to emerge. For  SOCs with $\Omega\ll \omega_{ho}$, it is sufficient to focus on the original degenerate ground state manifold  (denote as ${\cal M}$), which consists of all spin configurations. The particle distribution is fixed by the Slater Determinant $D$, which is a Fermi sea of the lowest $N$ states of the  harmonic trap, $D=\sum_{P}
(-1)^{P}u_{0}(x_{P1})u_{1}(x_{P2})...u_{N-1}(x_{PN})$. 
It can be  shown that \begin{equation}
D(x_1, x_2, ..., x_N) = C\prod_{N\geq i>j \geq 1}(x_i - x_j) e^{-\sum_{i=1}^{N} x_{i}^{2}/2a_{ho}^2}, \label{D}
\end{equation}
where $C$ is the normalization constant and $a_{ho}=(M\omega_{ho})^{-1/2}$. As particle number increases, the density profile of Eq.(\ref{D}) approaches that of the free Fermi gas given by Thomas-Fermi approximation. 

To obtain the ground state in the presence of SOC, one diagonalizes $V^{SO}$ within ${\cal M}$. This can be done using the 
angular momentum  basis $\chi^{(S, M)}$. However, 
  the construction of these basis and  the evaluation of the matrix elements $\langle SM | V^{(SO)} |S'M'\rangle$ are very complicated, (see Ref. \cite{CH}).  Here, we introduce a ``spin-ordered" basis,  which simplifies the calculation considerably and allows one to obtain an effective Hamiltonian which makes the underlying physics very  transparent.  In a  spin-ordered state  $|\xi_1, \xi_2, ... \xi_N \rangle\equiv |\vec{\xi}\rangle$, one encounters  a sequence  of spins $\xi_{1}$, $\xi_{2}$, ..$\xi_{N}$  as one moves from left to right.  Precisely, $|\xi\rangle$ is defined as 
\begin{eqnarray}
&&\langle x_1,...x_N; \mu_1,...\mu_N | \vec{\xi} \rangle = \frac{1}{\sqrt{N!}} D(x_1, x_2, ..., x_N)
\nonumber  \\
&& \hspace{0.4in} \sum_{P} 
 \theta(x_{P_1}, x_{P2}, .., x_{P_N}) 
\prod_i 
 \delta_{\xi_i,\mu_{P_i}} \label{spin-order},
\end{eqnarray}
where $P$ is a permutation of the integers $(1,2, .., N)$, and 
\begin{eqnarray}
\theta(x_{P_1}, x_{P2}, .., x_{P_N}) & =1 \,\,\,\,\,  {\rm if}  \,\,\,\,\,  x_{P_1}< x_{P2}...<x_{P_N} \nonumber  \\
   & = 0 \,\, \,\,\,\,  {\rm otherwise}.  \hspace{0.7in}
\end{eqnarray}
It is straightforward to verify that $ \{ |\vec{\xi}\rangle \}$ is an orthonormal basis,  $\langle \vec{\xi}'| \vec{\xi}\rangle= \prod_{i=1}^{N} \delta_{\xi_{i}', \xi_{i}}$.

Evaluating $V^{(SO)}=\sum_{i=1}^{N}V^{(SO)}_{i}$ in the degenerate manifold ${\cal M}$, 
we have
$V^{(SO)} = \sum_{\vec{\xi'},\vec{\xi}} |\vec{\xi'}\rangle \langle \vec{\xi} |  V^{(SO)}_{\xi', \xi}$,
\begin{eqnarray}
 V^{(SO)}_{\xi', \xi}   =  -   \Omega \sum_{i=1}^{N} b_{i} \delta_{\xi_{1}', \xi_{1}} \delta_{\xi_{2}', \xi_{2}}...
(S_{-})_{\xi_{i}', \xi_{i}} ...  \delta_{\xi_{N}', \xi_{N}}    \hspace{0.3in}  \label{M-element}  \\
b_{i} = |b_{i}|e^{i\alpha_{i}} =   \int {\rm d}{\bf x} |D|^2 \theta(x_{1},...,x_{i},..., x_{N}) e^{iqx_{i}}    \hspace{0.3in}  \label{b}
\end{eqnarray}
where ${\rm d}{\bf x} = {\rm d}x_1  {\rm d}x_2 ...  {\rm d}x_N$.  
Eq.(\ref{M-element}) is precisely the matrix element of the following simple hamiltonian 
 \begin{eqnarray}
H_{\rm eff} & = -  \Omega \sum_{i=1}^{N}  (b_{i}S_{i}^{-}  + h.c.) = - \Omega\sum_{i=1}^{N} {\bf b}_{i}\cdot{\bf S}_{i} 
\hspace{0.3in}   \label{Heff} \\
{\bf b}_{i} & = |b_{i}| (\hat{\bf x}  {\rm cos}\alpha_{i} + \hat{\bf y}  {\rm sin}\alpha_{i} ).  \hspace{1.4in}
\label{bi} \end{eqnarray}
In other words, the spin-ordered basis enables one to recast the original hamiltonian into that of a set of independent spins. In this formulation,  the positions of the fermions are absent from the problem, and their effects are  succinctly included in the set of effective magnetic fields $\{ {\bf b}_{i} \}$ acting on the ordered spins. Once the ground state of $H_{\rm eff}$ is obtained, (which is of the form $\sum_{ \{ \xi_{i} \}} C_{\vec{\xi}} |\vec{\xi}\rangle$),  the ground state wavefunction of $H$ in real space can also be obtained using Eq.(\ref{spin-order}).

{\it (C) Ground state spin spiral at $g=+\infty$}:  
Noting that the ground state of the $i$-th spin in Eq.(\ref{Heff}) is
 $|G\rangle_{i} = \sum_{\mu_{i}=\pm 1} e^{-i\alpha_{i}\mu_{i}/2}|\mu_{i}\rangle$, which depends on the direction (i.e. the angle $\alpha_{i}$) but not the magnitude of ${\bf b}_{i}$. The ground state of Eq.(\ref{Heff}) and its energy are then 
\begin{equation}
|G\rangle = \sum_{ \{ \xi_{i} \} } e^{ - i\sum_{i} \alpha_{i} \xi_{i} /2} |\vec{\xi}\rangle, \,\,\,\,\,  E_{G} = - \Omega \sum_{i=1}^{N} |b_{i}|.
\label{G}  \end{equation}
To understand the spin structure of this ground state, let us first consider the ``sequential" particle density, 
\begin{eqnarray}
n_{i}(x) &= \int {\rm d}{\bf  x} |D|^2 \theta(1,2, ..i, ..N) \delta(x-x_{i}), \,\,\,\, \,\,    \label{ni}  
\end{eqnarray}
and we have $ \int dx n_i(x)=1$. Eq.(\ref{ni}) describes the density of the $i$-th particle one encounters as one moves from left to right\cite{density}.  It is easy to see that 
\begin{eqnarray}
n(x) = \sum_{i=1}^{N}n_{i}(x), \,\, \, {\bf S}(x) = \sum_{i=1}^{N}{\bf s}_{i}(x), \label{nxsx}   \\
{\bf s}_{i}(x) = \frac{1}{2} n_{i} (x)\hat{\bf b}_{i}, \,\,\,\, \,  \hat{\bf b}_{i} =  \hat{\bf x} {\rm cos} \alpha_i +  \hat{\bf y} {\rm sin} \alpha_i.   \label{si}
\end{eqnarray}
Eq.(\ref{b}) shows that the complex ``magnetic field" $b_{i}$ is simply the Fourier transform of $n_{i}(x)$,  
\begin{eqnarray}
b_{i} =   \tilde{n}_{i}(q). \label{bi}
\end{eqnarray}

We shall write the spin vector as ${\bf S}(x)= S(x) ( {\rm cos}\alpha(x) , {\rm sin}\alpha(x) )$. 
 If the spins were classical, they would  follow the magnetic field ${\bf B}(x)$, and the spin vector would take the classical form 
${\bf S}_{c}(x) = (n(x)/2)({\rm cos}qx, {\rm sin}qx)$. However, due to quantum mechanical motion of the fermions, 
the angle $\alpha(x)$ and the magnitude $S(x)$ can be different from $qx$ and $n(x)/2$.

Finally, we note that the total spin of the system is 
\begin{eqnarray} 
\langle G| {\bf S}^2 |G\rangle = \frac{1}{2}N+\frac{1}{4}\left| \sum_{i=1}^{N} e^{i\alpha_{i}} \right|^2.  \label{S2}
\end{eqnarray}
When  all $\alpha_i=0$,  (i.e. when ${\bf B}$ is uniform or $q=0$), we have ${\bf S}^2 = (N/2+1)N/2$ and the system is a full ferromagnet pointing to $\hat{\bf x}$. For non-zero $q$, $\alpha_{i}\neq 0$ and  ${\bf S}^2$ decreases rapidly with $q$.   

\begin{widetext}

\begin{figure}[hbtp]
\includegraphics[height=4.7cm,width=18.7cm]{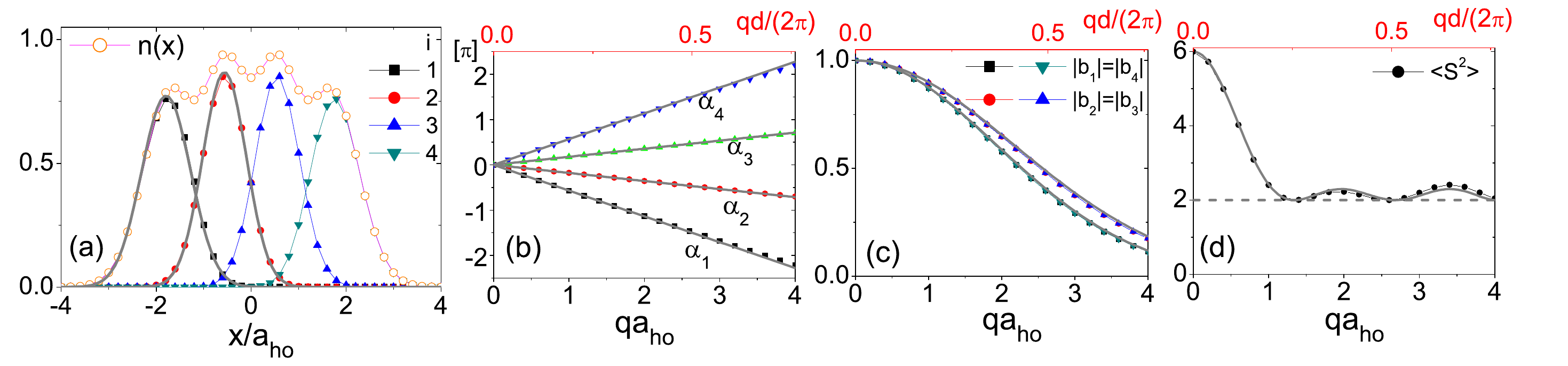}
\caption{(Color Online). (a) Total density $n(x)$ and the density of each ordered spin $n_i(x)$ in units of $1/a_{ho}$, (b)  the phase $\alpha_i$, (c) the magnitude $|b_i|$ of local effective field, and (d) the total spin of ground state $\langle {\bf S}^2\rangle$ for a $N=4$ system. The solid curves in (a) are Gaussian-fit with width $\sigma_1/a_{ho}=0.73, \ 
\sigma_2/a_{ho}=0.65$ and center location ${\bar x}_1/a_{ho}=-1.788,\  {\bar x}_2/a_{ho}=-0.56$ (see text). Due to reflection symmetry of the trapping potential, ${\bar x}_{1}= - {\bar x}_{4}$,  ${\bar x}_{2}= - {\bar x}_{3}$. 
The quantity $d$ in  the upper x-axis of (b), (c), (d) is the average inter-particle spacing $d=\frac{1}{N-1} \sum_{i=1}^{N} ({\bar x}_{i+1}-{\bar x}_i)$. The solid curves in (b), (c), and (d) are given by Eq.(\ref{bi_gauss}) and (\ref{S2_2}).  In (d), $\langle {\bf S}^2\rangle$ drops from $(N/2+2)N/2$ at $q=0$ to $ N/2$ around $qd/(2\pi)\sim 1/N$. } \label{fig1}
\end{figure}

\end{widetext}

\begin{figure}[hbtp]
\includegraphics[height=11cm,width=8.8cm]{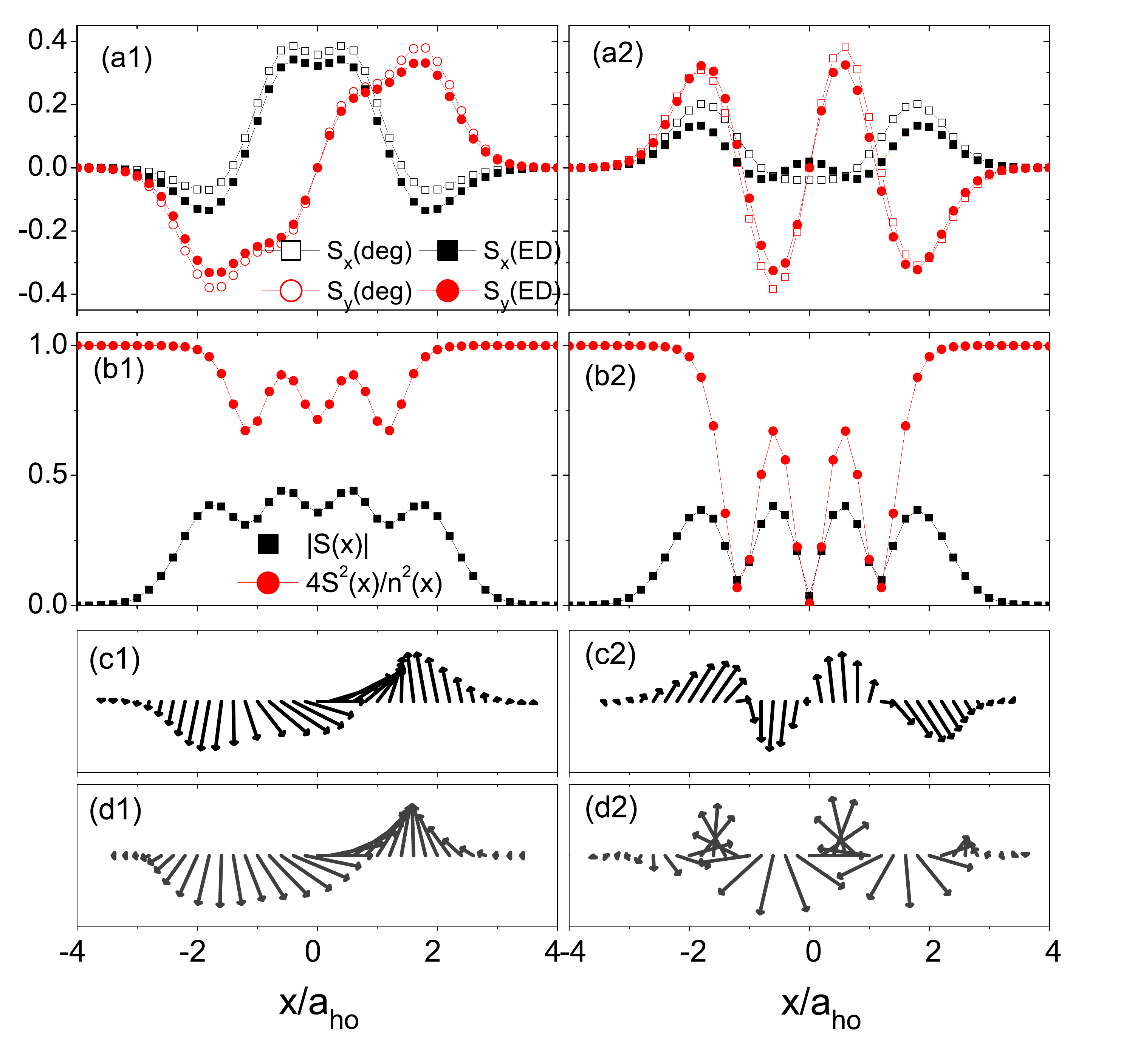}
\caption{(Color online). Spin density profiles for four particles at different $q$. The left (right) panel corresponds to $qa_{ho}=1\ (3)$, or $qd/(2\pi)=0.2 \ (0.6)$. (a$1^0$, a$2^0$): Spin density ${\bf S}(x)=(S_x,S_y)$ at $\Omega/\omega_{ho}=0.1$ for non-interacting case. 
(a1,a2):   Spin density at strong repulsion limit $g/(a_{ho}\omega_{ho})=12$. 
Results of degenerate perturbation calculation and exact diagonalization are respectively given by hollow and solid marks.
(b1,b2): The magnitude of the spin density, and  its ratio with number density.  (c1,c2): Vector plot of ${\bf S}(x)$ in coordinate space.  For comparison, (d1,d2) shows the classical spin spiral  ${\bf S}_{c}(x)=\frac{n(x)}{2}(\cos qx, \sin qx)$ which is parallel to the local field ${\bf B(x)}$. } \label{fig2}
\end{figure}

We shall first present  the results for a four particle system using Eqs(5-17)\cite{cal}.  These results are of direct relevance to the experiments on fermion clusters\cite{Jochim}. They also indicate the general behavior of systems with large number of particles.
Fig.1(a) shows the sequential density $n_{i}(x)$ for a four fermion system at infinite repulsion. One sees that the fermions are almost equally spaced with separation $d$,  and each $n_{i}(x)$ has  a width  $\sigma_i$ comparable to inter-particle spacing. Our numerical results (dots in Fig.1(a)) show that $n_{i}(x)$ is well approximated by a Gaussian (the solid curves in Fig.1(a)), 
\begin{equation}
n_i(x)\rightarrow \frac{1}{\sqrt{\pi}\sigma_i} e^{-(x-{\bar x}_i)^2/\sigma_i^2},  \label{ni}
\end{equation} 
where ${\bar x}_i=\int x n_i(x)$ is the averaged location of the $i-$th particle, and $\sigma_i$ is   of order of interparticle spacing $d$ (See Fig.1a).

Consequently, from  Eq.(\ref{bi}) we have
\begin{equation}
\alpha_i=q{\bar x}_i,\ \ \ \ |b_i|=\exp(-q^2\sigma_i^2/4).   \label{bi_gauss}
\end{equation}
These results, shown as solid lines and curves in Fig.1(b) and (c), match well our exact numerical calculations (dots).  
The decrease of $|b_{i}|$  with increasing $q$  as shown in Eq.(\ref{bi_gauss}) is due to the quantum motion of the $i$th fermion about its average position ${\bar x}_{i}$, as each fermion samples all the magnetic fields ${\bf B}(x)$ in 
the interval ${\bar x}_{i-1}< x_{i} < {\bar x}_{i+1}$.  The more rapid is the rotation of the magnetic field  ${\bf B}(x)$ (i.e. larger $q$), the wider is its range of ${\bf B}$ being sampled, and the weaker is the  effective magnetic field ${\bf b}_{i}$. 
With the approximation $ \overline{ {\bar x}_{i+1}-{\bar x}_i} = d$ and the result $\alpha_{i}= q\bar{x}_{i}$ (Eq.(\ref{bi_gauss})), 
we find from  Eq.(\ref{S2}) that 
\begin{equation}
\langle{\bf S}^2\rangle = N/2 + [\sin(qNd/2)/\sin(qd/2)]^2/4.  \label{S2_2}
\end{equation}
$\langle {\bf S}^2 \rangle$ drops quickly from $\sim N^2$ to $\sim N$ as $q$ increases from zero to $2\pi/(Nd)$, 
where the field rotates one round across the  sample. 
Again, the analytic approximation Eq.(\ref{S2_2}) (shown as solid curve in Fig.1(d)) matches well with the exact numerical results (dots).

Fig.2 shows the details of the spin density profile ${\bf S}(x)$. To illustrate the development of the spin spiral, we have studied its evolution with $g$ by exact diagonalization. For non-interacting case, we show in Fig.2(a1$^0$) and (a2$^0$) the spin densities with a weak SOC ($\Omega=0.1 \omega_{ho})$ for  different $q$.  The magnitudes of $S_x,\ S_y$ for both cases are very small (of the order of $\Omega/\omega_{ho}$). However, as $g$ exceeds $g_{c}\sim \Omega^{-1}$ (see more details below), the magnitudes of $S_x,\ S_y$ quickly rise to their  $g=+\infty$ values (given by Eqs(14,15)), as shown 
  in Fig.2(a1,a2). 

Fig.2(b1) and (b2) show the spin amplitude $|{\bf S}(x)|$ for different $q$. For small $q$, Fig.2(c1) and (d1) show that ${\bf S}(x)$ is close to a  classical spiral ${\bf S}_{c}(x)$. However, when  $qd\sim \pi$,  where $d$ is the interparticle spacing, ${\bf S}(x)$ deviates significantly from ${\bf S}_{c}(x)$, as seen from Fig.2(c2) and (d2). Moreover, the magnitude of ${\bf S}(x)$ varies strongly with position,  breaking up into chunks that track
 the locations of the fermions (see Fig.2(b2)).  This is because 
the spin density ${\bf S}(x)$ within the interval  $\overline{x}_{i}$ and $\overline{x}_{i+1}$ is essentially $(n_{i}(x) \hat{\bf b}_{i}+ n_{i+1}(x) \hat{\bf b}_{i+1})/2$. It depends on the widths $\sigma_{i}$ ($\sigma_{i}\sim d$) and  the direction of ${\bf b}_{i}$. The former reflects the quantum fluctuation of a fermion about  its  averaged position. The latter is the direction of the magnetic field at $\overline{x}_{i}$. 
When $qd \sim \pi$, we have ${\bf b}_{i+1} \sim - {\bf b}_{i}$. The Gaussians interfere destructively, leading to a very small spin density between $\overline{x}_{i}$ and $\overline{x}_{i+1}$, and large mis-alignment between the spin density ${\bf S}(x)$ and the local field ${\bf B}(x)$. 



Finally, to understand the critical repulsion $g_{c}$ for the emergence of the large spin spiral,  we recall that  in the absence of SOC, the energy of the singlet ground state near infinite interaction behaves as $E(1/g) -E(0) = -NC/g$, where $C=d(E/N)/d(-1/g)|_{g=\infty}$ is a positive constant\cite{adiabatic}. On the other hand, the energy gain from SOC is $-N \Omega e^{-q^2\sigma^2/4}$.  The spin spiral state will be more stable than the singlet state if $1/g <1/g_c\equiv \Omega e^{-q^2\sigma^2/4}/C$, which is always satisfied for sufficiently strong repulsion.

In summary, we have pointed out the large response of  a 1D spin-1/2 repulsive Fermi gas to a tiny amount of SOC, which only occurs when the repulsion is sufficiently strong. What triggers this  response is  the large spin degeneracy of the system at infinite repulsion. Since such  degeneracy also exists in other 1D systems such as  large spin bosons and fermions at infinite repulsion, the phenomena discussed here can also be found in these systems, probably in a richer forms of spin textures.  



XC acknowledges the support of NSFC under Grant No. 11104158, No. 11374177, and programs of Chinese Academy of Sciences. TLH acknowledges the support by DARPA under the Army Research Office Grant Nos. W911NF-07-1-0464, W911NF0710576.

\clearpage

\end{document}